\documentclass[12pt]{iopart}
\usepackage{amssymb}

\newcommand{\be}{\begin{equation}}
\newcommand{\ee}{\end{equation}}
\newcommand{\bea}{\begin{eqnarray}}
\newcommand{\eea}{\end{eqnarray}}

\newcommand{\AU}{\, \mathrm{AU}}
\newcommand{\D}{\mathcal{D}}
\newcommand{\GB}{\mathcal{G}}

\begin{document}

\title{Solar System Constraints on $f(\GB)$ Dark Energy}

\author{Stephen C Davis}

\address{
Laboratoire de Physique Theorique d'Orsay, 
B\^atiment 210, Universit\'e Paris-Sud 11, 91405 Orsay Cedex, France
}

\eads{\mailto{sdavis@lorentz.leidenuniv.nl}}

\begin{abstract}
Corrections to solar system gravity are derived for $f(\GB)$ gravity theories,
in which a function of the Gauss-Bonnet curvature term is added to the
gravitational action. Their effects on Newton's law, as felt by the planets,
and on the frequency shift of signals from the Cassini spacecraft, are both
determined. Despite the fact that the Gauss-Bonnet term is quadratic in
curvature, the resulting constraints are substantial. It is shown that they
practically rule out $f(\GB)$ as a natural explanation for the
late-time acceleration of the universe. Possible exceptions are when
$f(\GB)$ reduces to something very close to a cosmological constant, or
if the form of the function $f$ is exceptionally fine-tuned. 

\end{abstract}

\section{Introduction}

The current accelerated expansion of our universe cannot be explained by
conventional general relativity if our universe contains only standard
matter and radiation. Some form of additional dark energy, such as a
cosmological constant, may be the source of the acceleration, although such
models suffer from serious fine-tuning problems. It may be that the
acceleration instead comes from corrections to Einstein gravity. Such an
approach has the potential to avoid the fine-tuning problems, resulting in a
far more credible theory. One candidate for effective dark energy is
the quadratic curvature Gauss-Bonnet term 
\be
\GB = R^2 - 4R_{\mu\nu} R^{\mu\nu} + R_{\mu\nu\rho\sigma} R^{\mu\nu\rho\sigma}
\, ,
\ee
which is a natural addition to the Einstein-Hilbert action~\cite{Lovelock}.

In fact, on its own in four dimensions, the contribution of the Gauss-Bonnet
term to the gravitational field equation is trivial. For it to
have an effect the theory needs to have extra dimensions, such as in the
brane world scenario with one~\cite{BW} of more~\cite{BW2} additional
dimensions, or alternatively the Gauss-Bonnet term can be coupled to a scalar
field. Another related possibility is to add a function $f(\GB)$ of it to the
gravitational action~\cite{fG}. Such a theory will be the subject of this
article.

Its potential to give a more elegant explanation for the universe's accelerated
expansion makes gravity modification very appealing. However it does have one
major drawback, namely that deviations from general relativity will be felt on
all scales, not just cosmological ones. In particular gravity will be altered
within our solar system, where high precision tests of general relativity have
been performed. As we will show for $f(\GB)$ gravity, the corrections
are generically too large, and allow it to be ruled out as
a solution to the dark energy problem (with some severely fine-tuned
exceptions).

Within our solar system, gravitational fields are weak. Since the Gauss-Bonnet
term is quadratic in curvature, it might be expected that its effects will be
sub-dominant, and not significantly constrained. However as we will
demonstrate, this reasoning is flawed. In section~\ref{sec:cos} we will show,
after introducing the theory, that if the cosmological contribution of
$f(\GB)$ is to be large enough to act as dark energy, the couplings in $f$
must be extremely large. This will greatly magnify the effects of $\GB$ in the
solar system, and produce corrections to the Newtonian and post-Newtonian
potentials, which we will derive in section~\ref{sec:ss}. Strong constraints
on these corrections arise from planetary motion and light bending
measurements, which are analysed for a more general theory in
section~\ref{sec:con}. In section~\ref{sec:die} these are applied to $f(\GB)$
gravity, and it is shown that the gravitational effects within the solar
system are indeed large enough to conflict with the observational data that is
available, and thus allow this dark energy candidate to be practically ruled
out.

\section{Dark energy from $f(\GB)$ gravity}
\label{sec:cos}

Working in units with $c=1$, we will be studying the theory
\be
\mathcal{L} = \sqrt{-g}\left[\frac{R}{2} +f(\GB)\right] \, .
\label{L1}
\ee
It is equivalent to coupling a Gauss-Bonnet term to a scalar
field with a potential, but without an explicit kinetic term (in contrast to
the Gauss-Bonnet theories studied in~\cite{us, gilles}). In particular, if we
take the action
\be
\mathcal{L} = 
 \frac{1}{2}\sqrt{-g}\left[R 
+ \xi(\phi) \GB - 2V(\phi) \right]
\label{L2}
\ee
with
\be
V(\phi)= -f(\phi) + \phi f'(\phi) \, , \qquad \xi(\phi)=2f'(\phi) \, ,
\ee
and then vary it with respect to the scalar field, we obtain 
$\phi = \GB$. Substituting this back into the action~\eref{L2}, it reduces to
the theory~\eref{L1}.

Varying the action~\eref{L1} with respect to $g_{\mu\nu}$, we obtain
the gravitational field equations 
\bea 
&& G_{\mu\nu}
+8(R_{\mu\rho\nu\sigma} + 2 R_{\rho[\nu} g_{\sigma]\mu} 
- 2 R_{\mu[\nu} g_{\sigma]\rho} 
+ R g_{\mu[\nu}g_{\sigma]\rho}) \nabla^\rho \nabla^\sigma f'(\GB)
\nonumber \\ && \hspace{2.5in} {}
+  [\GB f'(\GB) -f(\GB) ] g_{\mu\nu}
= 8\pi G_0 T_{\mu\nu} \, .
\eea
These can also be obtained from~\eref{L2}. The constant $G_0$ in the
above equation is the gravitational coupling of matter, which may or
may not be the same as the gravitational coupling $G$ that we perceive
on Earth.

In a cosmological background ($ds^2 =- dt^2 +a(t)^2 dx_3^2$), we find
$\GB  = 24 H^2 (\dot H + H^2)$, with $H = \dot a/a$.
The Friedmann equation can be written as
\be
1 = \frac{8 \pi G_0}{3 H^2}\rho_{\rm mat} + \Omega_\GB \, ,
\label{Fried}
\ee
where the Gauss-Bonnet density fraction is
\be
\Omega_\GB \ = \ -8H \partial_t f'(\GB) +  \frac{\GB f'(\GB)-f(\GB)}{3H^2}
\, . \label{OGBa}
\ee
We will focus on the case $f(\GB) = C \GB^n$. The expression~\eref{OGBa}
then reduces to
\be
\Omega_{\GB} =  \frac{C (n-1)}{3}[-12 (1+3w)]^n 
\left[1 - 12 n \frac{(1+w)}{1+3w}\right] H^{2(2n-1)} \, ,
\label{OGB}
\ee
where $w = -1 -2\dot H/(3H^2)$ is the effective equation of state for the
universe. An important point to note is that if $\Omega_{\GB} \sim 0.7$, as is
required if $f(\GB)$ is to give a sufficient contribution to the dark energy
density, then $C \sim H_0^{2-4n}$. This is extremely large (for $n>1/2$), and
so we see that the Gauss-Bonnet term must be very strongly coupled if it is to
have any chance of explaining the accelerated expansion of our universe.

Note that if a solution of the above theory is solve the dark energy problem,
$\Omega_\GB \sim 0.7$ is necessary, but not sufficient. We also need the
solution to actually produce enough acceleration (so $w \approx -1$), and for
the cosmological evolution of our universe to reach it (after passing though a
period of matter domination, exactly like the one that occurred in our
universe). In this work we will mainly be concerned with the magnitude of
$\Omega_\GB$, which will be enough to rule out most possible $f(\GB)$ dark
energy models.

\section{Solar system gravity}
\label{sec:ss}

Within the solar system we can use the approximate Post-Newtonian
metric~\cite{will}
\be
ds^2= -(1+2 \Phi) \, dt^2  + (1-2 \Psi) \, \delta_{ij} dx^i dx^j 
+ \Or(\epsilon^{3/2}) \, .
\label{metric}
\ee
with $\Phi, \Psi \sim \epsilon$ and $\dot \Phi, \dot \Psi \sim
\epsilon^{3/2}$. The gravitational fields of the sun and planets are
relatively weak and slowly varying, and $\epsilon \lesssim 10^{-5}$. This
allows us to study gravity with a perturbative expansion in
$\epsilon$. Note that the metric~\eref{metric} is motivated by the properties
of our solar system, and not by the choice of gravitational
theory.

Making no assumptions about the size of $f$ or its derivatives, we
find that to leading order in $\epsilon$, the gravitational field equations are
\be \fl
\Delta \Phi = 4 \pi G_0 \rho_{\rm mat}
+ f(\GB) - \GB f'(\GB)-4\D(f'(\GB), \Phi+\Psi) 
+ \epsilon^2\Or(1, \GB f'', \GB ^2 f''') + \epsilon \, \Or(f,\GB f') 
\label{Ph1}
\ee
\be \fl
\Delta \Psi = 4 \pi G_0 \rho_{\rm mat}
+\frac{\GB f'(\GB)- f(\GB)}{2} -4 \D(f'(\GB), \Psi)
+ \epsilon^2 \Or(1, \GB f'', \GB^2 f''') + \epsilon \, \Or(f,\GB f') 
\label{Ps1}
\ee
and $\GB = 8 \D(\Phi,\Psi) + \Or(\epsilon^3)$. We have introduced the operators
\be
\Delta F = \sum_i F_{,ii} \,  , \qquad
\D(X,Y) = \sum_{i,j} X_{,ij} Y_{,ij} - \Delta X \Delta Y \, ,
\ee
which for functions with only $r$ dependence reduce to
\be
\Delta F = \frac{1}{r^2}\partial_r(r^2 \partial_r F) \,  , \qquad
\D(X,Y) = -\frac{2}{r^2}\partial_r(r \partial_r X \partial_r Y) \, .
\ee
In principle, the higher order (in $\epsilon$) terms of the above
expansion could give significant contributions if their coefficients
are very large. However for $f=C \GB^n$, the corresponding
coefficients are all of comparable size, and the higher terms can be dropped. 
Furthermore when $f=C \GB^n$, we see that the $f(\GB)$ and $\GB
f'(\GB)$ terms are higher order in $\epsilon$ than the $\D(\cdots)$
terms. For the rest of this section we will assume this form for the
theory, and so can drop those two terms from the field equations.
In this work we will be mainly interested in $n>0$, since theories with
negative $n$ have been studied elsewhere in the literature. In
particular, it has been shown while they satisfy solar system
constraints~\cite{karel1}, they are unable to solve the dark energy
problem~\cite{sepa}. We will discuss these points in more detail in
section~\ref{sec:die}.

The usual, Einstein gravity ($f\equiv 0$) solution is $\Phi = \Psi = -U$,
with
\be 
U = 4 \pi G_0  \int d^3 x' 
\frac{\rho_{\rm mat}(\vec x',t)}{|\vec x- \vec x'|} \, .
\ee
Treating the sun as a uniform sphere we find that $U$, and the Gauss-Bonnet
term, reduce to 
\be
U_{\rm ext} = \frac{G_0 m_\odot}{r}
\, , \qquad
\GB_{\rm ext} \approx 48 \frac{(G_0 m_\odot)^2}{r^6}
\ee
outside the sun ($r > R_\odot$), and
\be
U_{\rm int} = \frac{G_0 m_\odot}{2 R_\odot}\left(3-\frac{r^2}{R_\odot^2}\right)
\, , \qquad
\GB_{\rm int} \approx -48\frac{(G_0 m_\odot)^2}{R_\odot^6}
\label{in}
\ee
inside it ($r < R_\odot$).

A general analysis for non-trivial $f$ would be difficult. However we know
that for any viable theory, the resulting gravitational potentials must
be very close to the standard form. Following the approach
of~\cite{us}, we start by assuming that the potentials are very close
to the usual $1/r$ form. Deviations from this can be treated
perturbatively. By bounding their size, we can derive constraints on $f$.
Of course for a wide range of $f(\GB)$ theories, $\Phi$ and $\Psi$
will be radically different from $1/r$, and our approach will not give
their approximate form correctly. However, since we already know (by
definition) that such theories fail to give the observed gravitational
potentials, and there is no need to study them in the first place. Hence
our approach will cover all potentially relevant cases.

For the interior of the sun, we see that 
$\D(f'(\GB_{\rm int}), U_{\rm int})=0$. The interior solution~\eref{in}, up to
the addition of a constant, is therefore still valid (recall that we are
ignoring the $f$, $\GB f'$ terms as they are sub-dominant for $f=C \GB^n$).

For the exterior solution we take $\Phi = -Gm_\odot/r + \delta \Phi$ and 
$\Psi = -Gm_\odot/r + \delta \Psi$. Here $G$ is the approximate gravitational
coupling that we perceive, and need not be equal to the bare coupling $G_0$.
The perturbations to the usual potentials resulting from the $f$ dependent
$\D(\cdots)$ operator are given by $\delta \Phi$ and $\delta \Psi$. Such an
analysis will only be valid if $\delta \Phi, \delta \Psi \ll U$, which will
indeed be the case for us. The field equations \eref{Ph1} and \eref{Ps1}
reduce to
\be
\Delta \, \delta \Phi \approx 8 \D\left(C n\GB^{n-1}, \frac{Gm_\odot}{r}\right)
\, , \qquad
\Delta \, \delta \Psi \approx 4 \D\left(C n\GB^{n-1}, \frac{Gm_\odot}{r}\right)
\ee
with $\GB \approx 48 (Gm_\odot)^2/r^6$. Solving these, and requiring
continuity of $\Phi$, $\Psi$, and their first derivatives, gives the perturbed
exterior potentials 
\bea \fl
\ \Phi = -\frac{m_\odot}{r}\left[G_1 - \frac{GA}{(1+s) \, r^{s}}\right] 
 \, , \qquad
&\Psi =  -\frac{m_\odot}{r}\left[G_1 - \frac{GB}{(1+s) \, r^{s}}\right]
 & \quad \ \ ({\rm for} \ s \neq -1)
\nonumber \\ \fl
\ \Phi = -\frac{m_\odot}{r}\left[G_1 - G A \, r \ln r\right]  \, , 
\qquad
&\Psi =  -\frac{m_\odot}{r}\left[G_1 - G B \, r \ln r \right]
 & \quad \ \ ({\rm for} \ s = -1) 
\label{PhiPsi}
\eea
where we have introduced
\be \fl
A = 2 B = 2 n (n-1) \, C 48^n  (G m_\odot)^{2(n-1)}
\, , \qquad G_1 = G_0 + A R_\odot^{-s} 
\, , \qquad s = 2(3n-2) \, . \label{Phip}
\ee
We see that there are finite width effects appearing in the gravitational
potential. This is  due to the presence of higher than second order
derivative operators appearing in the gravitational field equations. This also
implies that the usual treatment of the sun, and other objects, as
point-like masses is no longer valid.

Clearly the above expressions~\eref{PhiPsi} are different from the usual
Einstein gravity result, and are likely to come into conflict with the high
precision measurements of gravity in the solar system. By bounding the
corrections, we will be able to constrain the parameters in the function
$f(\GB)$, and hence its dark energy contribution.

\section{Solar system gravity tests}
\label{sec:con}

We will now review two sources of gravitational constraints coming from the
solar system. First a test of Newton's law from planetary motion, and then
the frequency shift of light rays, which tests a
relativistic effect. The results will apply to any theory giving potentials of
the form~\eref{PhiPsi}, and not just $f(\GB) = C \GB^n$ gravity. We will apply
them to the $f(\GB)$ dark energy models in section~\ref{sec:die}.

Corrections to the Newtonian potential can be bounded by considering their
effect on bodies orbiting the sun. For an expression of the
form~\eref{PhiPsi}, the gravitational acceleration experienced by a body at
distance $r$ from the sun is
\be
g_{\rm acc}(r) = -\frac{d \Phi}{dr} =-\frac{m_\odot}{r^2}
\left[G_1 - A \frac{G}{r^{s}}\right] \equiv \frac{G m_{\rm eff}(r)}{r^2}
\, . \label{gacc}
\ee
For a body following an elliptical orbit with semi-major axis $a$, 
Kepler's third law gives the period of the orbit as
$2 \pi \sqrt{a^3/(G m_\odot)}$.  Bounds on corrections to the effective
mass are then related to the uncertainties in the measurement of $a$ by
\be
\frac{\delta m_{\rm eff}(a_\alpha)}{m_\odot} <
3 \frac{\delta a_\alpha}{a_\alpha} \, , 
\label{mcon}
\ee
where the index $\alpha$ runs over all bodies which are orbiting
the sun.  Values of $\delta a$ for the planets can be found
in~\cite{pitjeva}. The above relation has previously been used to bound dark
matter in the solar system~\cite{anderson}, the cosmological
constant~\cite{SSlambda}, and another class of Einstein-Gauss-Bonnet gravity
models~\cite{us}.

As it stands, the above constraint~\eref{mcon} depends on $G_1$, which is
related to the unmeasured constant $G_0$. By combining constraints from two
bodies, we can eliminate it to obtain
\be
\left|A \left(a^{-s}_\alpha -a^{-s}_\beta\right)\right|
 < 3 \left( \frac{\delta a_\alpha}{a_\alpha} 
+ \frac{\delta a_\beta}{a_\beta}\right) \, ,
\label{Pcon}
\ee
which needs to be satisfied for all choices of $\alpha, \beta$. 
For $s\geq-1$ the strongest bounds come from the inner planets, firstly
because these are the bodies for which we have the best data, and secondly
since they are closest to the sun, where the gravity corrections are
strongest.

To satisfy the bound~\eref{Pcon}, either $A$ will need to be very small, $s$
will need to be close to zero (in which case the corrections to the potential
will mimic Newtonian gravity), or $s$ will need to be less that $-1$ (in which
case the corrections are suppressed at the smaller distances where better data
is available).

Further constraints come from signals between man-made
spacecraft and the Earth. The sun's gravitational field produces a time delay
in the signals, measurement of which provides an additional test of gravity in
the solar system. Furthermore, unlike the planetary constraint~\eref{Pcon},
this is sensitive to relativistic effects, and so can be used to rule out
models which mimic Newtonian gravity.

For a light-ray starting at the Earth, passing close to the sun's
surface, continuing to the spacecraft ($r_e$ from the sun), and then returning
by the same route, the time delay in the signal is
\be
\Delta t = -2\int_{-r_e}^{a_\oplus} [\Phi(r)+\Psi(r)] dx \, .
\ee
The signal's path is approximated by $r=\sqrt{x^2+b^2}$, where the impact
parameter $b$ is defined as the smallest value of $r$ on the light ray's
path. A small value of $b$ will maximise the above time delay. 
Particularly good data was obtained for the Cassini spacecraft while making its
journey to Saturn. During 2002, the impact parameter fell as low as
$b = 1.6 R_\odot \approx 0.0074 \AU$ (the value of $r_e$ at this point was
$8.43 \AU$). The actual measurements obtained were not of the time delay, but
its frequency shift given by~\cite{cassini}
\be
y = \frac{d \Delta t}{dt} 
\approx \frac{d \Delta t}{db} \frac{db}{dt} 
= -\frac{4 G m_\odot}{b}\frac{db}{dt}
\left[2 + (2.1\pm2.3) \times 10^{-5} \right]
\, . \label{cassy}
\ee

For potentials of the form~\eref{PhiPsi}, the frequency shift evaluates to
\be
y = -\frac{4m_\odot}{b}\frac{d b}{dt} \left( 
2 G_1 \mathcal{I}_0 - [A+B] G \mathcal{I}_s \right) \, ,
\label{y1}
\ee
where we have defined 
\be
\mathcal{I}_s = \frac{b^2}{2}
 \int_{a_\oplus}^{-r_e} \frac{dx}{(b^2+ x^2)^{(3+s)/2}} \, ,
\ee
which evaluates to a hypergeometric function. A similar analysis was applied
to a different Einstein-Gauss-Bonnet theory in~\cite{us}, with $s=6$ due to
the simplifying assumptions imposed on the model. 

For small impact parameter, $b \ll a_\oplus, r_e$, which is the case for the
above Cassini bound~\eref{cassy}, the leading order behaviour of the above
integral is
\be 
\mathcal{I}_s \approx \left\{ \begin{array}{cc} \displaystyle
\frac{\sqrt{\pi} \, \Gamma([2+s]/2)}{2 \, \Gamma([3+s]/2)}\frac{1}{b^{s}}
& s > -2 \vspace{0.05in} \\  \displaystyle
\frac{b^2}{2} \ln\frac{4r_e a_\oplus}{b^2} 
& s = -2 \vspace{0.05in} \\  \displaystyle
-\frac{b^2}{2(2+s)}\left[r_e^{-(s+2)}+a_\oplus^{-(s+2)}\right]
& s < -2 \end{array} \right.
\, .
\ee
In particular $\mathcal{I}_0 \approx 1$. 

Comparing~\eref{cassy} and~\eref{y1} gives a bound on $A+B$, although this is
of limited use, since like~\eref{gacc} it depends on the undetermined quantity
$G_1$. To eliminate that, we combine the constraint with~\eref{mcon}, giving
\be
-(2 \times 10^{-6}) -6\frac{\delta a_\alpha}{a_\alpha} 
< 2 A a_\alpha^{-s} - (A+B) \mathcal{I}_s 
 < (4.4  \times 10^{-5}) +6 \frac{\delta a_\alpha}{a_\alpha} \, .
\label{Ccon}
\ee
Note that the above constraint will need to be satisfied
for all choices of $\alpha$.

Another well-known solar system gravity test comes from the
perihelion precession of Mercury. However we will not consider it here as the
linearised analysis of section~\ref{sec:ss} is insufficient to determine it,
and a higher order (in $\epsilon$) expansion is needed. As it turns out, the
above results will be sufficient to eliminate $f(\GB)$ as a viable dark energy
component.

\section{Constraining $f(\GB)$ gravity}
\label{sec:die}

Combining the expression for the effective dark energy fraction~\eref{OGB}
with the planetary motion constraint~\eref{Pcon}, and taking the current
effective equation of state for our universe to be $w=-1$, we obtain the bound
\be
|\Omega_{\GB}| < \frac{H_0^{2(2n-1)} }{2^{n+1}n  r_g^{2(n-1)}
 |a_\alpha^{2(2-3n)} -a_\beta^{2(2-3n)}|}
\, \left( \frac{\delta a_\alpha}{a_\alpha} 
+ \frac{\delta a_\beta}{a_\beta}\right) \, ,
\label{Pcon2}
\ee
where $H_0^{-1} \approx 8.8 \times 10^{14} \AU$ is the Hubble distance, and 
$r_g = G m_\odot \approx 9.7 \times 10^{-9} \AU$ is the gravitation radius of
the sun. For $n>1$ we find the strongest constraint comes from taking 
$(\alpha, \beta)$ to be Mercury ($a \approx 0.39 \AU$, 
$\delta a/a \approx 1.8 \times 10^{-12}$) and the Earth ($a \approx 1 \AU$, 
$\delta a/a \approx 0.98 \times 10^{-12}$). This implies
\be
|\Omega_\GB| \lesssim 1.6 \times 10^{-43} \, ,
\ee
which is clearly far too small to account for the dark energy of our
universe. For smaller $n$, a significant dark energy fraction may be
possible. Using all the planets in our solar system, we find $\Omega_\GB
\lesssim 0.1$ unless
\be
\left|n-\frac{2}{3}\right| \lesssim 1.1 \times 10^{-27}
\qquad {\rm or} \qquad
n \lesssim 0.074 \, .
\label{except}
\ee
The two numerical values come respectively from the Earth-Mercury and
Earth-Uranus combinations (Uranus has $a \approx 19 \AU$, 
$\delta a/a \approx 1.3 \times 10^{-8}$).
Note that for the above bounds on $\Omega_\GB$ we assumed the equation of
state $w=-1$ for our universe. Other values will give slightly weaker or
stronger constraints, but they will be of the same order of magnitude.

Although the first of the exceptions~\eref{except} mimics the correct
Newtonian limit of gravity, it fails to produce the correct relativistic
effects. Using the Cassini constraint~\eref{Ccon},
we find that for the parameters~\eref{Phip},
\be \fl
-\frac{2 \times 10^{-6}}{6} -\frac{\delta a_\alpha}{a_\alpha} 
<  \frac{2^n n r_g^{2n-2}}{H_0^{4n-2}}
 (2a_\alpha^{4-6n} - 3\mathcal{I}_{6n-4}) \, \Omega_\GB
 < \frac{4.4  \times 10^{-5}}{6} + \frac{\delta a_\alpha}{a_\alpha} \, ,
\ee
where we have again used~\eref{OGB} with $w=-1$. Taking $\alpha$ to be
Mercury, we find that  $|\Omega_\GB| \lesssim 1.2 \times 10^{-20}$ for 
$n\geq 0.66$, which includes the first range in~\eref{except}, thus ruling it
out.

It should be pointed out that the errors $\delta a_\alpha$ appearing
in~\cite{pitjeva} are in fact the statistical errors in the values of
$a_\alpha$ coming from a least-squares fit of observations. The
real errors may well be an order or two of magnitude higher, and would
give correspondingly weaker bounds. However the above bounds are so
strong that even this would not be enough to save the theory.

The second exception~\eref{except} to the bound~\eref{Pcon2} includes the
parameter range $n <0$. Such models have been studied elsewhere, for
example~\cite{karel1,sepa,karel}.  We see that (for $n <1/2$) the correction
to the Newtonian potential for a general mass $M$ is  
\be
\delta \Phi \propto n\left(\frac{r^{3}}{G M}\right)^{1-2n} \, ,
\label{dPhiM}
\ee
which actually decreases for larger masses. Obviously, the above
expression is not valid at very large distances, where $\delta \Phi$
will cease to be a small perturbation. However for couplings of the
size needed to solve the dark energy problem, $C \sim H_0^{2-4n}$,
the above $\delta \Phi$ will remain small within the solar system.

We see that for a heavy object such as the sun the corrections to
$\Phi$ and $\Psi$ are suppressed, which is why (for low enough $n$)
they do not conflict with our solar system constraints. On the other
hand, we see that the above expression becomes large for small masses,
implying that the weak field approximation we have been using breaks
down. This suggests that the theory will predict non-Newtonian gravity
for table-top gravity experiments, and so will disagree with
observation. However it must not be forgotten that laboratory
experiments are performed on the Earth, which has a large
gravitational field of its own. The above expression~\eref{dPhiM} no
longer applies since it is now the Earth which gives the dominant
contribution to the Gauss-Bonnet term $\GB$. The resulting bounds on
the theory are then far weaker, and do not rule out dark energy models
with $n<0$~\cite{karel1,karel}.

Although $f(\GB)$ gravity with inverse powers of $\GB$ can give a large enough
$\Omega_\GB$, and avoids conflict with solar system gravity tests, this is
not enough to solve the dark energy problem. During its early evolution, our
universe passed through a matter dominated, decelerating phase (with 
$\GB =24 H^2 (\ddot a /a) < 0$), before entering the current accelerated phase
(with $\GB >0$). The Gauss-Bonnet term must therefore have passed though zero
at some point. We see from the cosmological field equations~\eref{Fried},
\eref{OGBa} that if $f(0)$ is not bounded, $\GB=0$ corresponds to curvature
singularity, and is unreachable (for finite $H$). Since no transition to
accelerated expansion occurs, inverse powers of $\GB$ do not give viable
dark energy models~\cite{sepa}. However, they may still be relevant to the dark
matter problem~\cite{karel}.  

So far we have only considered a special subset of the modified gravity
theory~\eref{L1} where the function $f(\GB)$ is some power of the Gauss-Bonnet
term. A more general theory can be expanded as a power series
\be
f(\GB) = \sum_n C_n \GB^n \, ,
\ee
where the $n$ are not necessarily integers. Now each term gives
a correction to the Newtonian potential $\Phi$ with a different $r$ dependence,
and so barring extreme fine-tuning of the $C_n$, we can constrain each
of them separately. Similarly for $\Psi$. Hence, using the above
results, we find that for each of their contributions to the dark energy, 
$\Omega_\GB^{(n)} \ll 1$. The only exception is $0 \leq n \lesssim 0.074$, 
which is very similar to a cosmological constant.

Finally, we note that our gravity corrections have turned out to
be tiny, at least for $f(\GB) = C \GB^n$ theories with positive $n$,
and which also satisfy the solar system constraints. Hence the
leading order corrections $\delta \Phi$ and $\delta \Psi$ used in
section~\ref{sec:ss} are much smaller than $Gm_\odot/r$, and the
perturbative analysis used there is indeed self-consistent.  In the
case of $n=0$, the leading order non-zero corrections to $\Phi$ and
$\Psi$ are not covered by our analysis. This case is the same as a
cosmological constant, and has been covered elsewhere in the
literature (e.g.~\cite{SSlambda}). It is worth pointing out that a
very wide range of $f(\GB)$ theories do not have $\Phi\approx \Psi
\approx -G m_\odot/r$ as an approximate solution, and so the analysis
of this paper does not apply to them. However, since the gravitational
field of the sun in these theories has no resemblance to what we have
observed in the solar system, they are already ruled out, and there is
no need to apply this paper's analysis.

\section{Conclusions}

We have seen that a careful analysis of solar system gravity provides a
powerful probe of higher curvature gravity theories, and can be used rule
out modified gravity models which are intended to solve the dark energy
problem. Specifically, strong constraints on Newton's law are obtained from
the motion of the planets. These are sufficient to rule out a large fraction
of $f(\GB)$ dark energy models. One class of model that survives these
constraints are those with inverse powers of the Gauss-Bonnet term $\GB$,
although their cosmological evolution is not consistent with our universe.

The remaining $f(\GB)$ models which survive the Newtonian constraints either
mimic Newtonian gravity, or are very close to a cosmological
constant. Further strong constraints come from the measurement of the
frequency shift of signals form the Cassini spacecraft. These rule out the
models which mimic Newtonian gravity. Hence $f(\GB)$ is practically ruled out,
as it is only viable when it is virtually identical to a cosmological constant.

In principle, with enough fine-tuning, a model could be constructed where the
deviations from general relativity cancel for each of the planets (and for the
Cassini bound). The non-standard gravity effects could then still be large
enough to account for the dark energy. However for this to work the
corrections must cancel for every choice of position $r$, and mass, for
which there is a planet, moon, asteroid, laboratory experiment, time delay or
frequency shift measurement. Even then, there is no guarantee that the
non-standard cosmological evolution will be accelerating, be free from
ghosts~\cite{sepa}, and satisfy all other cosmological tests, such as
producing correct growth of density perturbations, which is also a
problem for $f(\GB)$ gravity~\cite{fgcos}. The problem of ghosts could
even arise within the solar system, giving another way to constrain
the theory there. The presence of ghosts can be determined by
analysing perturbations of the solar system metric. However we will
not pursue this for our $f=C \GB^n$ dark energy ($n>0$)
solutions, since they have already been ruled out. Furthermore, we
did not actually obtain a solution to the field equations within
the solar system (expect when $\Omega_\GB$ is negligible), so we do
not have a background solution whose perturbations are worth analysing.

Following the example of $f(R)$
gravity~\cite{fR}, an $f(\GB)$ that reduces to constant for large $\GB$ and
zero at small $\GB$ may give a viable dark energy model while
satisfying solar system constraints. In fact any $f(\GB)$ whose form gives
acceptable corrections for solar system curvatures (e.g.\ a inverse
power of $\GB$), but produces acceleration at cosmological scales
could be viable. However such a theory will require even more
fine-tuning than a conventional cosmological constant, so it is
debatable whether this is an improvement.

Our strong constraints on $f(\GB)$ dark energy rely on the fact that we can
directly link its local and cosmological properties. This is
possible because the behaviour of the theory is fully determined by the form of
the metric (which is known at both scales) and the constant parameters which
determine $f$. As an alternative type of theory we could have extended the
action~\eref{L2}, to include additional kinetic terms for the
scalar $\phi$, as in~\cite{us, gilles}. These can all have
$\phi$-dependent couplings, and can include higher order terms as well
as the usual $(\partial_\mu \phi)^2$. The resulting theory
has the same number of degrees of freedom as the one studied in this
paper, suggesting the same analysis can be applied. However, the value
of the field $\phi$ is no longer directly determined by the metric:
$\phi =\GB$. Instead it is related to $g_{\mu\nu}$ by a differential
equation, whose solution can include integration constants which are
determined by cosmological rather than local scales. In particular, the
value of $\phi$ at the edge of the solar system is (roughly) zero for
$f(\GB)$ gravity, and arbitrary for the more general theories. Hence
in the latter case connecting the strong local constraints and
the cosmological evolution will require further analysis of the
theory's behaviour on intermediate scales. The theories studied
in~\cite{us, gilles} therefore still have the potential to solve the dark
energy problem, although of course further work is required. Another,
more obvious advantage of the above scalar-tensor theories, is
that they contain more coupling functions than the single one in
$f(\GB)$. With a greater range of theories to choose from, there is
more hope of finding one which satisfies both local and cosmological
constraints.

\ack
I am grateful to Christos Charmousis, Antonio De Felice, Kazuya Koyama,
Antonio Padilla, Karel Van Acoleyen, and the anonymous referee for
useful comments and discussions, and to the Netherlands Organisation for
Scientific Research (NWO) for financial support.


\section*{References}

\begin{thebibliography}{19}
\bibitem{Lovelock}
  D.~Lovelock,
  {\em The Einstein Tensor And Its Generalizations,}
  J.\ Math.\ Phys.\ {\bf 12}, 498 (1971)
  %%CITATION = JMAPA,12,498;%%
\bibitem{BW}  
  C.~Charmousis and J.~F.~Dufaux,
  {\em General Gauss-Bonnet brane cosmology,}
  Class.\ Quant.\ Grav.\  {\bf 19}, 4671 (2002)  [hep-th/0202107]
  %%CITATION = CQGRD,19,4671;%%
\nonum
  S.~C.~Davis,
  {\em Generalised Israel junction conditions for a Gauss-Bonnet brane world},
  Phys.\ Rev.\ D {\bf 67}, 024030 (2003)  [hep-th/0208205]
  %%CITATION = PHRVA,D67,024030;%%
\nonum
  E.~Gravanis and S.~Willison,
  {\em Israel conditions for the Gauss-Bonnet theory and the Friedmann 
  equation on the brane universe},
  Phys.\ Lett.\ B {\bf 562}, 118 (2003) [hep-th/0209076]
  %%CITATION = PHLTA,B562,118;%%
\bibitem{BW2}
  P.~Bostock, R.~Gregory, I.~Navarro and J.~Santiago,
  {\em Einstein gravity on the codimension 2 brane?,}
  Phys.\ Rev.\ Lett.\  {\bf 92}, 221601 (2004)  [hep-th/0311074]
  %%CITATION = PRLTA,92,221601;%%
\bibitem{fG}
S.~Nojiri and S.~D.~Odintsov,
  {\em Modified Gauss-Bonnet theory as gravitational alternative for dark
   energy,} 
  Phys.\ Lett.\  B {\bf 631}, 1 (2005)  [hep-th/0508049]
  %%CITATION = PHLTA,B631,1;%%
\bibitem{us}
L.~Amendola, C.~Charmousis and S.~C.~Davis,
  {\em Solar System Constraints on Gauss-Bonnet Mediated Dark Energy,}
  JCAP {\bf 0710}, 004 (2007)  [0704.0175 [astro-ph]]
  %%CITATION = JCAPA,0710,004;%%
\bibitem{gilles} 
G.~Esposito-Farese, 
  {\em Scalar-tensor theories and cosmology and tests of a
   quintessence-Gauss-Bonnet coupling,}
  gr-qc/0306018
%%CITATION = GR-QC/0306018;%%
\nonum 
G.~Esposito-Farese, 
   {\em Tests of scalar-tensor gravity,}
   AIP Conf.\ Proc.\ {\bf 736}, 35 (2004) [gr-qc/0409081]
%%CITATION = APCPC,736,35;%%
\bibitem{will}
C.~M.~Will, {\em The confrontation between general relativity and experiment,}
  gr-qc/0510072
  %%CITATION = GR-QC/0510072;%%
\nonum 
C.~M.~Will, {\em Theory and experiment in gravitational
physics}, Cambridge University Press
\nonum
  G.~Esposito-Farese,
  {\em Tests of Alternative Theories of Gravity,}
Proceedings of 33rd SLAC Summer Institute on Particle Physics (SSI
2005): Gravity in the Quantum World and the Cosmos, Menlo Park,
California, 25 Jul -- 5 Aug 2005, pp T025 
  %%CITATION = ECONF,C0507252,T025;%% 
\bibitem{karel1}
I.~Navarro and K.~Van Acoleyen,
  {\em On the Newtonian limit of Generalized Modified Gravity Models,}
  Phys.\ Lett.\  B {\bf 622}, 1  (2005)  [gr-qc/0506096]
  %%CITATION = PHLTA,B622,1;%%
\bibitem{sepa}
A.~De Felice and M.~Hindmarsh,
  {\em Unsuccessful cosmology with Modified Gravity Models,}
  JCAP {\bf 0706}, 028 (2007)
  [0705.3375 [astro-ph]]
%%CITATION = JCAPA,0706,028;%%
\bibitem{pitjeva}
E.~V.~Pitjeva, 
{\em High-Precision Ephemerides of Planets--EPM and Determination of 
Some Astronomical Constants,}
Solar System Research {\bf 39}, 176 (2005)
\bibitem{anderson}
J.~D.~Anderson, E.~L.~Lau, A.~H.~Taylor, D.~A.~Dicus
D.~C.~Teplitz and V.~L.~Teplitz
  {\em Bounds on Dark Matter in  Solar Orbit,}
  Astrophys.\ J. {\bf 342}, 539 (1989) 
  %%CITATION = DOE-ER40200-143;%%
\bibitem{SSlambda} 
M.~Sereno and P.~Jetzer, 
  {\em Solar and stellar system tests of the cosmological constant,}
  Phys.\ Rev.\ D {\bf 73}, 063004 (2006) [astro-ph/0602438]
  %%CITATION = PHRVA,D73,063004;%%
\bibitem{cassini}
  B.~Bertotti, L.~Iess and P.~Tortora,
  {\em A test of general relativity using radio links with the Cassini
  spacecraft,}
  Nature {\bf 425}, 374 (2003)
  %%CITATION = NATUA,425,374;%%
\bibitem{karel}
  I.~Navarro and K.~Van Acoleyen,
  {\em Modified gravity, dark energy and MOND,}
  JCAP {\bf 0609}, 006 (2006)  [gr-qc/0512109]
  %%CITATION = JCAPA,0609,006;%%
\nonum
I.~Navarro and K.~Van Acoleyen,
  {\em Dark energy, MOND and sub-millimeter tests of gravity,}
  astro-ph/0605322
  %%CITATION = ASTRO-PH/0605322;%%
\bibitem{fgcos}
  B.~Li, J.~D.~Barrow and D.~F.~Mota,
  {\em The Cosmology of Modified Gauss-Bonnet Gravity,}
  Phys.\ Rev.\  D {\bf 76}, 044027 (2007)  [0705.3795 [gr-qc]]
  %%CITATION = PHRVA,D76,044027;%%
\bibitem{fR}
 W.~Hu and I.~Sawicki,
  {\em Models of f(R) Cosmic Acceleration that Evade Solar-System Tests,}
  Phys.\ Rev.\  D {\bf 76}, 064004 (2007)
  [0705.1158 [astro-ph]]
  %%CITATION = PHRVA,D76,064004;%%
\nonum
S.~A.~Appleby and R.~A.~Battye,
  {\em Do consistent $F(R)$ models mimic General Relativity plus $\Lambda$?,}
  Phys.\ Lett.\  B {\bf 654}, 7 (2007)
  [0705.3199 [astro-ph]]
%%CITATION = PHLTA,B654,7;%%
\nonum
A.~A.~Starobinsky,
  {\em Disappearing cosmological constant in f(R) gravity,}
  JETP Lett.\  {\bf 86}, 157 (2007)
  [0706.2041 [astro-ph]]
  %%CITATION = JTPLA,86,157;%%
\end{thebibliography}
\end{document}